\documentclass[12pt]{iopart}
\usepackage{graphicx}
\usepackage{amssymb}
\usepackage{cite}

\begin{document}

\title[EIT with magnetically-induced $\Delta F=0, m_F=0 \rightarrow m_F=0$ probe transition]{Electromagnetically induced transparency with magnetically-induced $\Delta F=0, m_F=0 \rightarrow m_F=0$ probe transition}

\author{Armen Sargsyan, David Sarkisyan, Aram Papoyan}

\address{Institute for Physical Research, NAS of Armenia, Ashtarak-2, 0204 Armenia}
\ead{armen\_sargsyan@ipr.sci.am}
\vspace{10pt}
\begin{indented}
\item[]20 February 2024
\end{indented}

\begin{abstract}
Interest in magnetically induced (MI) transitions of alkali metal atoms is caused by the fact that their intensities can exceed the intensities of “regular” atomic transitions in a wide range of magnetic field (200 -- 4000 G). The goal of this work was to form and study, for the first time, an electromagnetically induced transparency (EIT) resonance in a strong magnetic field using a probe radiation tuned to $|F_g = F, m_F=0 \rangle \rightarrow |F_e = F, m_F=0\rangle$ MI transition, which is forbidden in zero magnetic field. Two narrow-band linearly-polarized cw diode lasers were used to form a EIT resonance on a $\Lambda$-type system of Cs atomic D$_2$ line in a strong transverse magnetic field (up to 1000 G). The resonance was formed in Cs atomic vapor nanocell with the atomic vapor column thickness of 850 nm.

\end{abstract}
%
%

\section{Introduction}

In recent years, magnetically induced (MI) transitions of alkali metal atoms, forming a large class of one hundred frequency-shifted transitions with interesting and important features, have attracted great interest, primarily due to the fact that in wide ranges of magnetic fields (200 -- 4000 G depending on particular atoms) their intensities can significantly exceed the intensities of regular atomic transitions \cite{1,2,3,4,5,6}. Particularly, electromagnetic-induced transparency (EIT) exploiting MI transitions enables formation of narrow resonances in an extended frequency range (the shift from regular transitions can reach 20 -- 30 GHz) \cite{7,8,9}.

One can distinguish two types of MI transitions, MI1 and MI2. Using atomic state representation in the form $|F, m_F \rangle$, where $F$ is the total momentum of atom, and $m_F$ is its projection, MI1 stands for transitions between the ground $F_g$ and excited $F_e$ levels $|F_g, 0 \rangle \rightarrow |F_e = F_g, 0'\rangle$ (here and after the prime notes upper level), which are forbidden in zero magnetic field, but experience a gigantic intensity increase with the of magnetic field. For $B >> B_0$ ($B_0$ is the magnitude of magnetic field induction $B_0 = A_{hfs} / \mu_B$, where $A_{hfs}$ -- magnetic dipole constant of the ground level of an atom, $\mu_B$ -- Bohr magneton), increasing intensity of these asymptotically approaches to a constant value (see curve $1$ in Fig. \ref{fig:Fig1}b) \cite{11,12}. For $^{133}$Cs, $^{85}$Rb and $^{87}$Rb atoms, $B_0 =$ 1700 G, 700 G and 2400 G, respectively.

Transitions between the ground and excited levels of the hyperfine structure with $F_e-F_g= \Delta F = \pm 2$ induced by an external magnetic fields are MI2 transitions. These transitions have been successfully exploited to form EIT resonances when atomic vapor is exposed to magnetic field \cite{13,14,15}. Nevertheless, contrary to the case of MI1, the intensity of MI2 transitions drops practically down to zero in a strong magnetic field $B >> B_0$. For these $B$-fields, EIT with MI1 transitions can be more appropriate.

In this paper we report the first experimental realization of electromagnetically-induced transparency in Cs atomic vapor involving MI1 transitions induced by a strong transverse magnetic field (up to 1000 G). We also discuss possible applications of the formed EIT resonances.

\section{Theoretical consideration}

\begin{figure}[h]
	\centering
	\begin{center}
		\includegraphics[width=380pt]{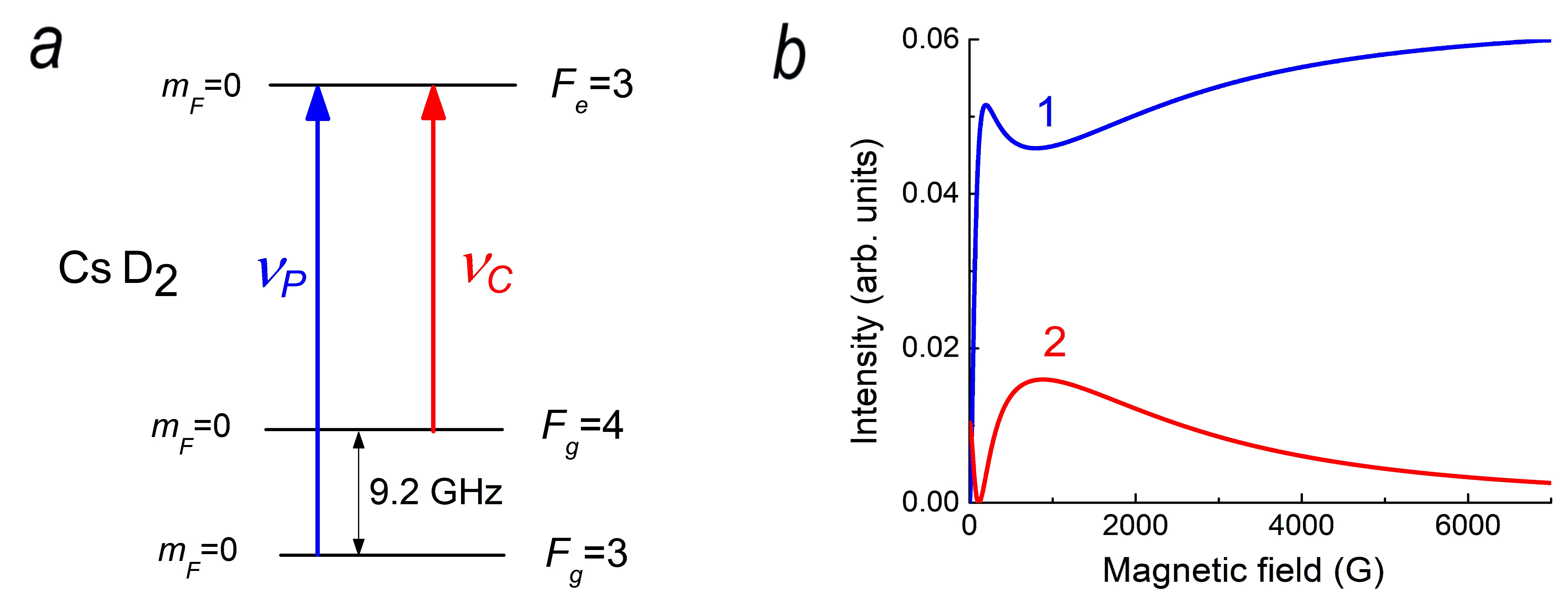}
		\caption{\label{fig:Fig1} a) Diagram of the $\Lambda$-system formed on Cs D$_2$ line with probe radiation on $|3,0 \rangle \rightarrow |3',0' \rangle$ MI1 transition and coupling radiation on $|4,0 \rangle \rightarrow |3',0' \rangle$ transition. b) Curves $\it1$ and $\it2$-- intensities of the probe (MI1) and coupling transitions, respectively, for the linearly $\pi$-polarized radiation versus magnetic field.}
	\end{center}
\end{figure}

The diagram of a $\Lambda$-system formed in the system of hyperfine levels of Cs atomic D$_2$ line involving MI1 transition is presented in Fig. \ref{fig:Fig1}a. Here the $|3,0 \rangle \rightarrow |3',0' \rangle$ MI1 transition serves as a probe, and the coupling radiation is tuned to $|4,0 \rangle \rightarrow |3',0' \rangle$ transition. Description of the theoretical model for calculation of the modification of the probability (intensity) of atomic transitions in a magnetic field using the Hamiltonian matrix, taking into account all transitions within the hyperfine structure, is presented in \cite{1,2,4}. 

Corresponding calculated dependences of the probe (MI1, curve $\it1$) and coupling (regular, curve $\it2$) transition intensities on magnetic field are shown in Fig. \ref{fig:Fig1}b for the case of linear $\pi$-polarized radiation. As is seen from the curve $\it1$, for $B >> B_0$ = 1700 G, the intensity of the MI1 transition increases, asymptotically approaching to a constant value, which makes it convenient for involvement in EIT process when atomic vapor is exposed to a strong magnetic field. As for the coupling transition (curve $\it2$), its intensity continuously declines within the same $B$-field range, yet being sufficient for realization of EIT. Note that the MI1 transition $\it1$ dominates over the regular transition $\it2$ for all magnetic induction values starting from $B \sim$ 200 G, which makes is attractive for practical applications.  

\section{Experiment} 

\subsection{Experimental setup}

\begin{figure}[h]
	\centering
	\begin{center}
		\includegraphics[width=300pt]{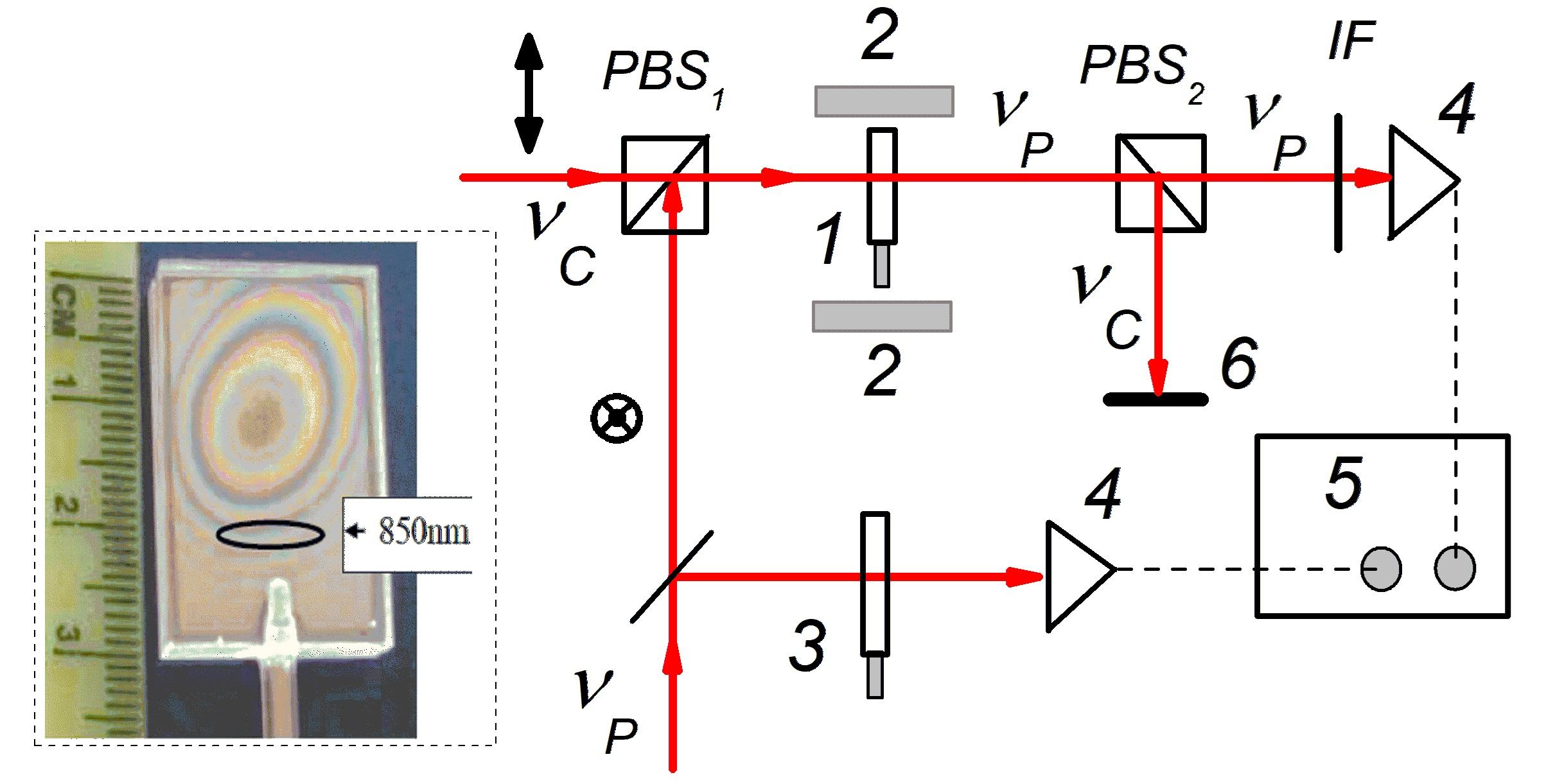}
		\caption{\label{fig:Fig2} The scheme of the experimental setup: $\it1$ -- Cs NC with the atomic vapor column thickness of $L=$ 852 nm; $\it{PBS_1}$, $\it{PBS_2}$ -- polarizing beam splitters; $\it2$ -- permanent magnets PM; $\it3$ -- auxiliary Cs NC to form a reference spectrum; $\it4$ -- photodiodes; $\it{IF}$ -- interference filter for $\lambda$ = 852 nm; $\it5$ -- oscilloscope Tektronix TDS2014B; $\it6$ -- beam damper. Left inset: photograph of the nanocell; the oval corresponds to the region where $L=\lambda \sim$ 850 nm.}
	\end{center}
\end{figure}

The scheme of the experimental setup is shown in Fig. \ref{fig:Fig2}. Radiation beams from two cw narrow-band (linewidth $\sim$ 1 MHz) tunable external cavity diode lasers (ECDL) with a wavelength $\lambda$ = 852 nm were used as sources of probe and coupling radiations \cite{16}. One of the ECDL (coupling laser) with a power $P_C$ = 10 -- 15 mW had a fixed frequency $\nu_C$, and the second one (probe laser) with a power $P_P$ = 0.1 -- 0.5 mW had a tunable frequency $\nu_P$. Linear polarizations of the coupling and probe lasers were purified by Glan prisms, and were mutually perpendicular (see $\it P$--marked double arrows in Fig. \ref{fig:Fig2}). The two radiation beams of $\varnothing \sim$ 1 mm were combined by a polarizing beam splitter $\it{PBS_1}$ and directed normally to a Cs atomic vapor nanocell (NC) \cite{17}. The EIT-resonance was formed in the Cs vapor column with thickness of $L= \lambda =$ 852 nm (the NC was placed in a non-magnetic oven). We should note that despite a small thickness and frequent damping collisions of atoms with the windows, the NC is convenient for formation of a EIT resonance in strong magnetic fields \cite{13}.

To record selectively the transmitted probe radiation spectrum bearing the EIT resonance without hindering contribution from the coupling beam, the latter was cut off by the $\it{PBS_2}$. Additional frequency selection of the probe radiation $\nu_P$ against the ambient light was attained by an interference filter IF ($\lambda$ = 852 nm, transmission width 10 nm). The transmitted probe radiation was recorded by a photodiode followed by an operational amplifier, the signal from which was fed to a digital oscilloscope. Part of the coupling laser radiation $\nu_C$ was branched to an atomic frequency locking system yielding a feedback "error" signal to stabilize its frequency (not shown in Fig. \ref{fig:Fig2}). 

The strong transversal magnetic field ($B$ is directed along the electric field $E$ of the probe radiation) was produced by a couple of ring-shaped permanent magnets PM made of neodymium-iron-boron alloy ($\varnothing$ = 60 mm, thickness $\sim$ 30 mm). The PMs were mounted on a non-magnetic translation stage allowing smooth variation between them. A convenient method for controlling and measuring a non-uniform magnetic field is given in \cite{18,19}. Part of the probe radiation $\nu_P$ was directed to an auxiliary Cs atomic vapor nanocell $\it3$ to form a reference transmission spectrum. 

\subsection{Experimental results and discussion}

\begin{figure}[h]
	\centering
	\begin{center}
		\includegraphics[width=440pt]{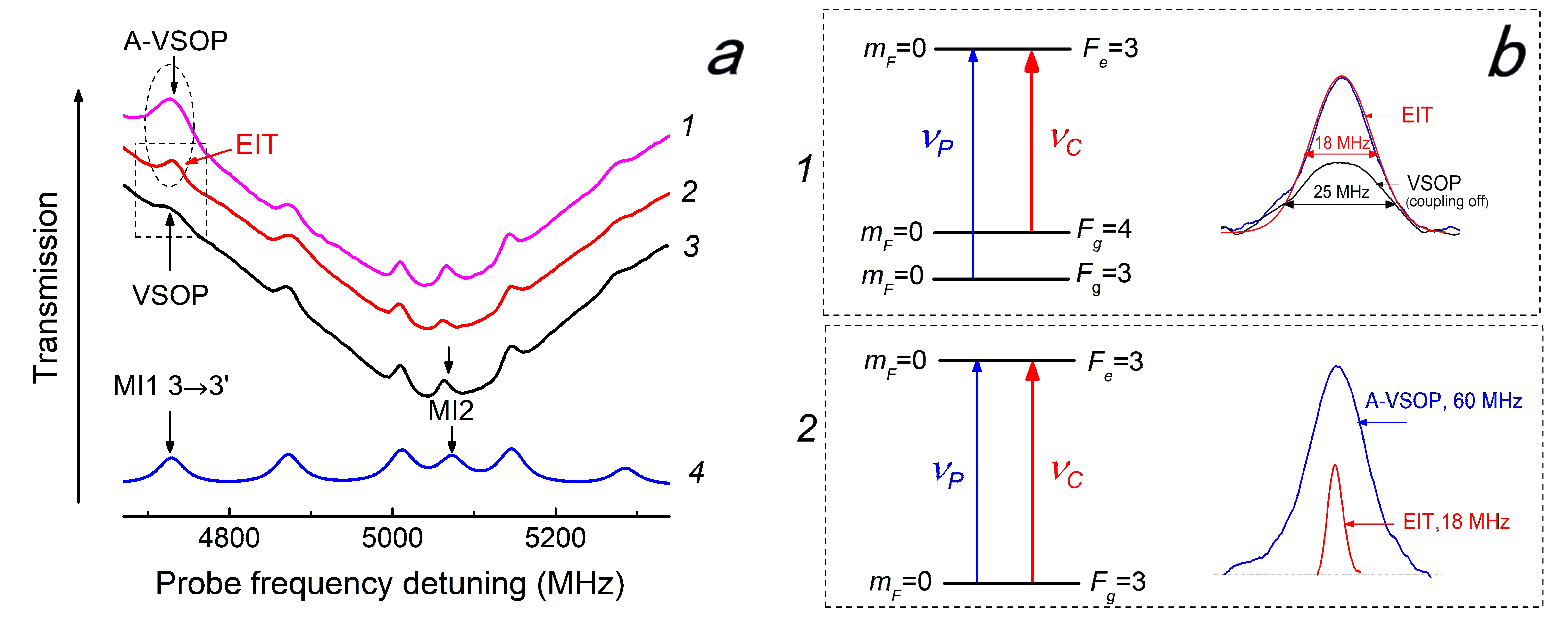}
		\caption{\label{fig:Fig3}a) Probe $\nu_P$ radiation transmission spectra: $\it{1}$ -- when $\nu_C$ coupling radiation is on $|3,0 \rangle \rightarrow |3',0' \rangle$ MI1 transition; $\it{2}$ -- when $\nu_C$ coupling radiation is on $|4,0 \rangle \rightarrow |3',0' \rangle$ regular transition; $\it{3,4}$ -- the experimental and theoretical transmission spectra of the probe only (the theory describes well frequency positions and amplitudes of VSOP resonances). b) Measurement configurations and resonance lineshapes (see text). General experimental conditions: Cs D$_2$ line, $B$ = 500 G ($\vec B$ $\parallel$ $\vec E$), NC thickness is $\sim$ 850 nm, NC temperature $\sim$ 100 $^\circ$C.}
	\end{center}
\end{figure}

In \cite{18} it was shown that "velocity-selective optical pumping" (VSOP) resonances are formed in the absorption spectrum of a nanocell with $L \sim \lambda$ thickness, when the laser radiation intensity is $>$ 1 mW/cm$^2$, showing a decrease in absorption. Spectral width of the VSOP resonances located at the frequencies of atomic transitions is 10 times narrower than the Doppler width. Curve $\it{1}$ In Fig. \ref{fig:Fig3}a shows the probe radiation transmission spectrum when $\nu_C$ is tuned to $|3,0 \rangle \rightarrow |3',0' \rangle$ MI1 transition (no $\Lambda$-system formed, see the diagram (2) in Fig. \ref{fig:Fig3}b). The measurement was done at $B$ = 500 G; the NC was heated to $\sim$ 100 $^\circ$C. One can see that one of the VSOP resonances on curve $\it{1}$, for which $\nu_P = \nu_C$, exhibits an obvious enhancement (so-called amplified, or A-VSOP). The full width at half-maximum (FWHM) of A-VSOP resonance is 60 MHz. The mechanism of A-VSOP formation is the following: the coupling radiation $\nu_C$ transfers some atoms from $F_g$=3 to $F_e$=3, followed by spontaneous decay to the level $F_g$=4, thus depleting the population of $F_g$=3. As a result, transmission of atomic vapor on 3 $\rightarrow$ $3'$ transition increases (optical pumping process \cite{10,22}). 

Curve $\it{2}$ in Fig. \ref{fig:Fig3}a presents the probe transmission spectrum when $\nu_C$ is tuned to $|4,0 \rangle \rightarrow |3',0' \rangle$ regular transition, under otherwise invariable conditions. In this case, a $\Lambda$-system formed (see the diagram (1) in Fig. \ref{fig:Fig3}b), and a EIT resonance appears when $\nu_P$ becomes equal to $|3,0 \rangle \rightarrow |3',0' \rangle$ MI1 transition frequency. The width of EIT resonance appearing at the same probe detuning value as for A-VSOP, is 18 MHz. Zoomed EIT and A-VSOP features are shown together for comparison in the right inset (2) of Fig. \ref{fig:Fig3}b. The EIT spectrum is approximated by a Gaussian envelope with a width of 18 MHz FWHM, and the VSOP resonance width is 25 MHz. The amplitude increase by a factor of $\sim$ 2 with a decrease in spectral width by a factor of 1.4 is the characteristic behavior of EIT resonances.

The contrast of EIT resonance, defined as a ratio of the peak EIT amplitude to the peak absorption of Cs vapor when the coupling beam is blocked, \cite{9} is $\sim$ 8 \%. It may seem surprising that EIT resonance with relatively high contrast is formed in the nanocell, as collisions of atoms with NC windows suppose to suppress coherency \cite{7,9,10}. But in the case of exact resonance of the coupling radiation frequency with corresponding transition (in our case $|4,0 \rangle \rightarrow |3',0' \rangle$), only atoms flying parallel to the nanocell windows, which practically do not experience wall collisions, contribute to formation of EIT \cite{8}. Once the coupling laser is detuned from the transition frequency, EIT formation is contributed by the atoms with non-zero velocity component perpendicular to the windows, which experience frequent collisions with the walls, thus suppressing the EIT process \cite{8}. Note that the presence of VSOP resonances in the NC spectrum makes it easy to precisely tune the coupling laser frequency to desired transition. This is among advantages of using the NC for EIT formation.

Curves $\it{3}$ and $\it{4}$ in Fig. \ref{fig:Fig3}a are the experimental and theoretical transmission spectra of the probe radiation (no coupling beam) at $B$ = 500 G, which exhibit both the $|3,0 \rangle \rightarrow |5',0' \rangle$ MI2 transition and $|3,0 \rangle \rightarrow |3',0' \rangle$ MI1 transition. 

The experimentally measured width of EIT resonance is consistent with the theoretical expectation. Indeed, the power-broadened EIT linewidth can be calculated by the expression given in \cite{10}:   

\begin{equation}
\gamma_{EIT} = \Omega_C^2 / \Gamma_{Dop} + \gamma_{23}, 
\end{equation}

\noindent where $\Omega_C$ is the Rabi frequency of the coupling radiation, $\Gamma_{Dop}$ is the one-photon Doppler width of Cs atoms $\sim$ 340 MHz, $\gamma_{23}$ is the coherence dephasing rate. For Rabi frequency, the estimate is obtained from $ \Omega_C/2 \pi = \Gamma_N (I/8)^{1/2}$ \cite{20}, where $I$ is the laser intensity in mW/cm$^2$, natural linewidth $\Gamma_N \approx$ 5.2 MHz, and thus for the coupling intensity $\sim$ 1000 mW/cm$^2$, $\Omega_C \approx$ 58 MHz. For the nanocell, $\gamma_{23} \sim$ 8 MHz \cite{21}, which yields $\gamma_{EIT} \approx$ 18 MHz.

\begin{figure}[h]
	\centering
	\begin{center}
		\includegraphics[width=250pt]{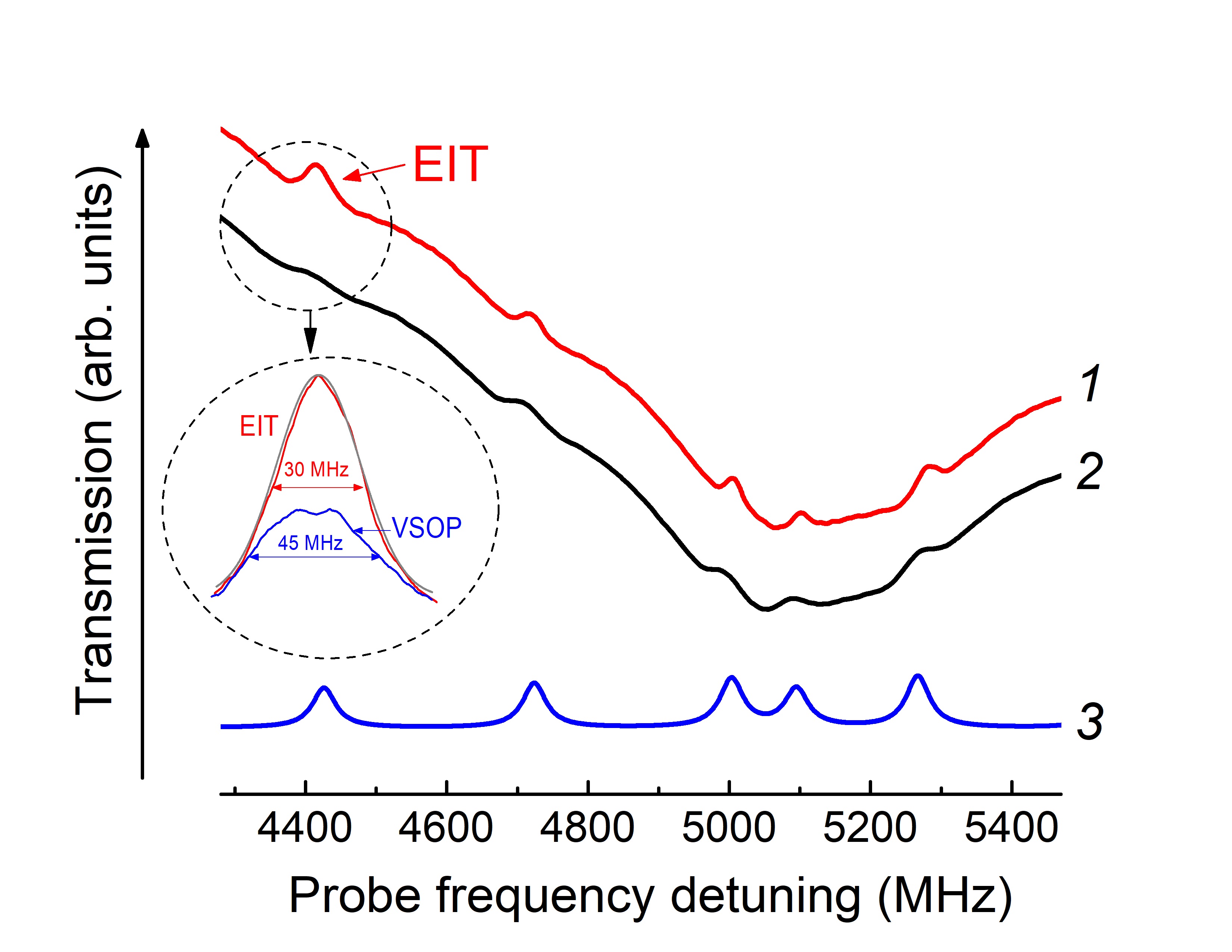}
		\caption{\label{fig:Fig4} Probe $\nu_P$ radiation transmission spectra: $\it{1}$ -- when $\nu_C$ coupling radiation is on $|4,0 \rangle \rightarrow |3',0' \rangle$ regular transition; $\it{2,3}$ -- the experimental and theoretical transmission spectra of the probe only. General experimental conditions: Cs D$_2$ line, $B$ = 1000 G ($\vec B$ $\parallel$ $\vec E$), NC thickness is $\sim$ 850 nm, NC temperature $\sim$ 100 $^\circ$C.}
	\end{center}
\end{figure}

Transmission spectra of the probe radiation for $B$ = 1000 G are presented in Fig. \ref{fig:Fig4}. Curve $\it{1}$ shows the probe transmission spectrum in the presence of $\nu_C$ coupling radiation tuned to $|4,0 \rangle \rightarrow |3',0' \rangle$ regular transition. EIT and VSOP resonances are clearly visible also in this case. These resonances are approximated by a Gaussian envelope (see the zoomed inset). The measured spectral linewidths are 30 MHz for EIT and 45 MHz for VSOP. The broadening of EIT and VSOP resonances is primarily due to the increase of transversal magnetic field inhomogeneity (magnets in this case are closer to the $\varnothing$ = 1 mm laser beam passing through the NC). Curves $\it{2}$ and $\it{3}$ are the experimental and theoretical spectra of the probe laser transmission in the absence of coupling radiation.
 
\begin{figure}[h]
	\centering
	\begin{center}
		\includegraphics[width=250pt]{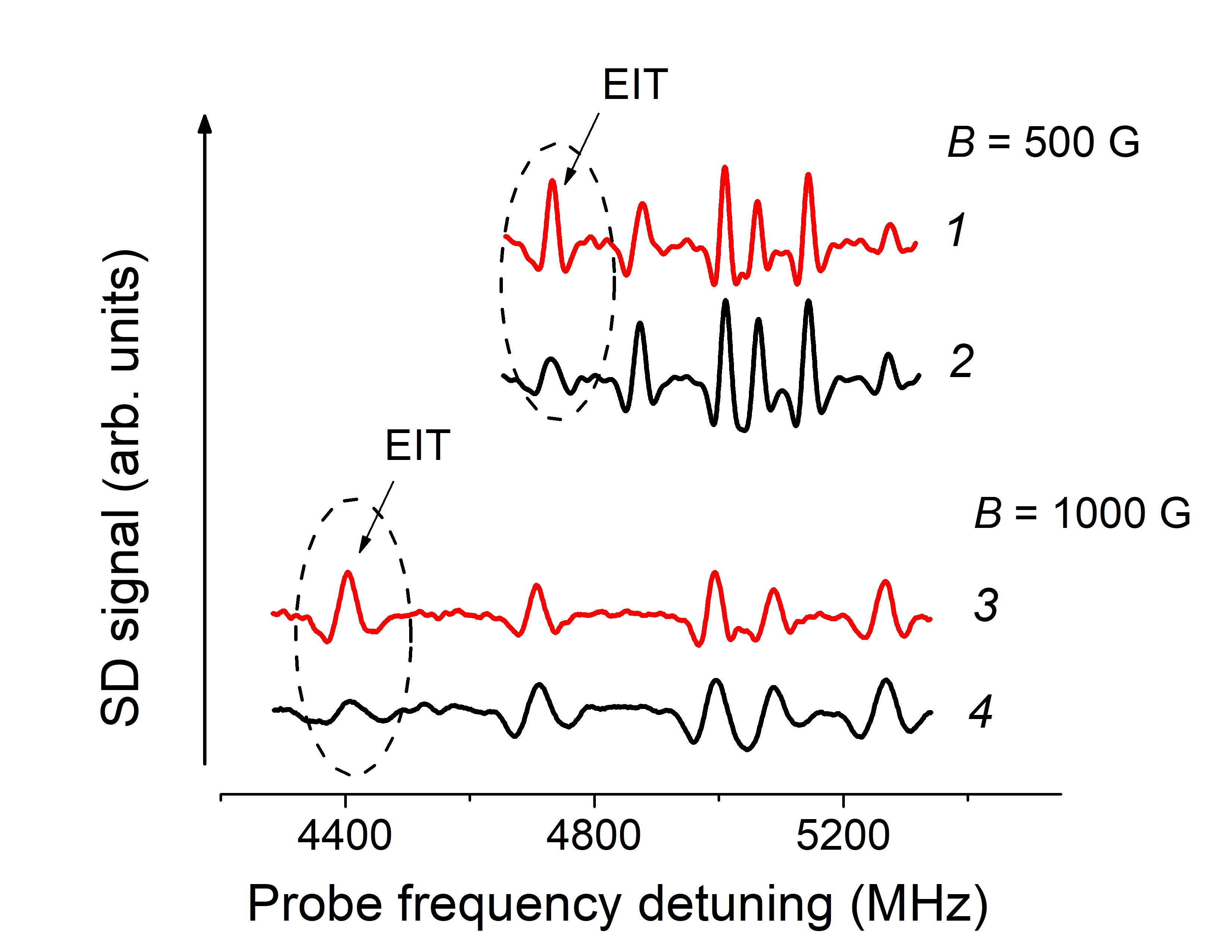}
		\caption{\label{fig:Fig5} SD transmission spectra of the probe radiation in the presence of coupling radiation (curves $\it{1}$ and $\it{3}$) and without the coupling radiation (curves $\it{2}$ and $\it{4}$) for $B$ = 500 and 1000 G.}
	\end{center}
\end{figure}

In \cite{23} it was shown that the second derivative (SD) of the recorded spectrum helps to reveal narrow, weak, and closely-spaced spectral features. Besides obtaining more pronounced resonance peaks or dips, this method also allows to make the overall slow variations of spectra horizontal. Figure \ref{fig:Fig5} shows the second derivatives of the probe radiation transmission spectra in the presence of coupling radiation (curves $\it{1}$ and $\it{3}$ for $B$ = 500 and 1000 G, respectively), and without the coupling radiation (curves $\it{2}$ and $\it{4}$ for $B$ = 500 and 1000 G, respectively). The corresponding raw spectra datasets were taken from spectra presented in Figs. \ref{fig:Fig3} and \ref{fig:Fig4}. It is clearly seen that thanks to SD processing, the narrow resonance peaks become more pronounced, and the overall Doppler-broadened absorption dip is outright straightened. 

As a final remark, we should note that EIT resonances with MI1 transitions can be obtained also for much stronger magnetic field. In \cite{24} it was shown that a $\Lambda$-system can be formed using one radiation with linear polarization and the other radiation with circular ($\sigma^+$ or $\sigma^-$) polarization. With a $\Lambda$-system formed by a linearly-polarized MI1 probe transition and correspondingly chosen strong regular circularly-polarized coupling transition it is possible to obtain a EIT resonance at $B \sim$ 7000 -- 9000 G (see curve $\it{1}$ in Fig. \ref{fig:Fig1}b).	

\section{Conclusion}

Summarizing, electromagnetically-induced transparency resonance was recorded in Cs atomic vapor interacting with two narrow-band linearly-polarized cw diode lasers and exposed to a strong transversal magnetic field, in a $\Lambda$-system formed within atomic D$_2$ line, where $|4,0 \rangle \rightarrow |3',0' \rangle$ transition was used for the coupling radiation, and the probe radiation, for the first time, was tuned to a MI1 magnetically-induced transition ($|3,0 \rangle \rightarrow |3',0' \rangle$). The EIT-resonance was observable for up to $B \sim$ 1000 G. Relatively high resonance contrast of 8 \% and the linewidth of 18 MHz was obtained for $B$ = 500 G.  

Note that EIT resonances with significantly narrower spectral width can be formed in centimeter-long atomic vapor cells with addition of a buffer gas, by using coherently-coupled probe and the coupling radiation derived from the same laser source \cite{9,10}. In this case, the coupling radiation intensity required for obtaining EIT resonance can be significantly lower, than in the present experiment.
 
Large number of magnetically-induced transitions in different alkali metal atoms and their isotopes (over 100 for D$_{1,2}$ lines), which become effective in strong magnetic field are of interest because their intensities can exceed intensities of regular atomic transitions in a wide ranges of magnetic field. Furthermore, their frequencies can be significantly shifted, thus allowing formation of narrow resonances in a wide range around regular atomic transitions. Electromagnetically induced transparency in strong magnetic field with $F_e-F_g = \Delta F =$ 2 MI2 type transitions between $F_g$ and $F_e$ levels has been successfully realized in \cite{10,11,12,13}. Nevertheless, for strong magnetic field $B >> B_0$, intensity of these transitions decreases asymptotically approaching to a zero. MI1 transitions $|F_g,0 \rangle \rightarrow |F_e=F_g,0' \rangle$ explored in present work preserve their high intensity at much higher values of $B$-field, extending the application areas.

\vspace{10pt}
The authors express their gratitude to A. Tonoyan for assistance in the work. The work was supported by the Higher Education and Science Committee of Armenia, in the frames of project N 1-6/IPR.

\end{document}